\documentclass[prl,twocolumn,nofootinbib,amsfonts]{revtex4}
\usepackage{graphicx,bm,amsmath,color}
\usepackage[latin1]{inputenc}

\newcommand{\be}{\begin{equation}}
\newcommand{\ee}{\end{equation}}
\newcommand{\comments}[1]{}

\begin{document}

\title{Constraints from Neutrino Decay on Superluminal Velocities}
\author{J.M. Carmona}
\email{jcarmona, cortes@unizar.es}
\affiliation{Departamento de F\'{\i}sica Te\'orica,
Universidad de Zaragoza, Zaragoza 50009, Spain}
\author{J.L. Cort\'es}
\email{jcarmona, cortes@unizar.es}
\affiliation{Departamento de F\'{\i}sica Te\'orica,
Universidad de Zaragoza, Zaragoza 50009, Spain}
\begin{abstract}
The splitting of neutrinos ($\nu \to \nu \nu {\bar \nu}$), a viable reaction for superluminal neutrinos, is shown to have phenomenologically relevant consequences if one accepts the recent OPERA results for the velocity of neutrinos. Neutrino splitting can be used to put strong constraints on the energy dependence of the velocity of propagation of neutrinos in a general analysis of modifications of relativistic kinematics and to propose observable effects due to the departures from special relativity in neutrino physics.
\end{abstract}

\maketitle

\section{Introduction}

Neutrinos are a special window to physics beyond the Standard Model~\cite{nu-new-physics}. They are the only particles which do not experience any long range interaction, which makes them particularly sensitive to possible new physics at the TeV scale or beyond. One example is the possible existence of large extra-dimensions~\cite{arkanietal} which could affect neutrinos but not the remaining elementary particles. Another example is the related property of neutrinos of being able to propagate freely over very large distances. This provides an amplification mechanism for effects coming from new physics.

The special role of neutrinos has been considered also in connection with possible departures from special relativity (SR) and
CPT invariance~\cite{tsukerman}. This can be understood if neutrinos have a special ability to explore the quantum nature of space-time, which is the more plausible origin for a departure from SR~\cite{qugraphen}. In this way, corrections to the relativistic kinematics for neutrinos have been considered previously as a possible alternative (or complement to masses) explanation of neutrino oscillations~\cite{nu-osc}. Very high-neutrino physics has also been suggested as a tool to examine quantum-gravity-induced corrections to the relativistic dispersion relation~\cite{VHE-nu-phys}.

Recently the OPERA neutrino experiment has reported~\cite{OPERA} the observation of a deviation in the time of flight of neutrinos with respect to SR corresponding to a relative difference of the muon neutrino velocity with respect to the speed of light
of\,\footnote{We use units where the speed of light $c$ is equal to one.}
$v-1 = (2.48 \pm 0.28\,\text{(stat.)} \pm 0.30\,\text{(sys.)}) \times 10^{-5}$.
This result, together with similar although less statistically significant indications in the same direction from another long base line experiment (MINOS)~\cite{MINOS}, comes out as a major surprise since the corrections to SR it implies are much larger than what one would expect if they came from the natural scale of quantum gravity~\cite{Flavio}.

What is more, this result is in contrast with the stringent limit of $v-1 < 2\times 10^{-9}$ set by the observation of (anti) neutrinos emitted by the SN1987A supernova~\cite{limitsSN}. Although we are dealing here with a different type of neutrinos (muon neutrinos in the accelerator experiments, versus electron neutrinos in the supernova case), flavor-dependent Lorentz violations are severely constrained by neutrino oscillations~\cite{Giudice}. However, neutrinos from SN1987A are in the 10\,MeV range, three orders of magnitude below the energy involved in the OPERA/MINOS experiments, which suggests the possibility that the apparent conflict between these observations might be due to an energy dependence of the deviations from SR, which is also natural from the point of view of the emergence of new physics controlled by a high-energy scale. All this provides a further motivation to consider the phenomenological consistency of superluminal neutrinos from a different perspective.

Among the observations which are sensitive to departures from SR a special role is played by those related to changes of thresholds and those related to the conversion of a forbidden reaction in SR into a viable reaction~\cite{Liberati,threshold}. In the case of neutrinos one has the possibility of neutrino decay according to reactions such as $\nu \to \nu \nu {\bar \nu}$ (neutrino splitting), $\nu \to \nu e^+ e^-$ (pair production), or $\nu \to \nu \gamma$ (photon emission). All these reactions are kinematically forbidden in SR, as one can easily see by examining them in the neutrino rest frame. However, they may be allowed if a neutrino with a given momentum has an energy greater than the SR energy as it is the case for superluminal neutrinos. The corresponding decay rate may be written as a function of the neutrino velocity, which allows to analyze the compatibility between the results obtained from OPERA and SN1987A for the neutrino velocity and the experimental observation (or non-observation) of the long distance propagation of neutrino fluxes of the corresponding energy.

In this work we will consider in detail the neutrino splitting, since it is a process which only involves the neutrino, and one does not need to make further assumptions on how the departures from SR affect other particles. Nevertheless, since the energies that we will consider are much larger than the electron mass, the electron pair production process will have a similar (order of magnitude) rate as the neutrino splitting process and the consequences and constraints that both type of reactions imply on the difference of the neutrino velocity with respect to the speed of light will also be similar. Finally, the photon emission process, being described by a loop diagram, is suppressed with respect to the other two and will give less severe constraints.

\section{Neutrino splitting width}

The splitting rate of a neutrino with a Planck scale modified dispersion relation
\be
E_\nu^2 \,=\, m_\nu^2 + p^2 + \eta \frac{p^4}{M_\text{Pl}^2}
\label{eta}
\ee
at tree level has been estimated in Ref.~\cite{Liberati}, yielding a decay width
\be
\Gamma_\nu \sim \frac{3 G_F^2}{64\pi^3}\, E_\nu^5 \left(\frac{\eta \,p^2}{M_\text{Pl}^2}\right)^4,
\ee
where the neutrino energy $E_\nu$ is well above the threshold of the process, so that the masses can be neglected in the calculation.

The result contains a factor of $3$ due to the three possible combinations of $\nu{\bar\nu}$ in the final state, a factor $G_F^2$ typical of weak processes, factors $(2\pi)$ characteristic of a three body decay, an $E_\nu^5$ factor that can be anticipated from dimensional analysis and the last factor involving the detailed form of the modification of the relativistic dispersion relation.
In fact the computation of the neutrino decay width does not depend on the explicit form of the modified dispersion relation~(\ref{eta}). In the general case one has
\be
\Gamma_\nu \sim \frac{3 G_F^2}{64\pi^3}\, E_\nu^5 \left(\frac{E^2-p^2}{p^2}\right)^4.
\label{Gamma0}
\ee

If one assumes the standard Hamiltonian relation $v (p)\,=\, \text{d} E/\text{d} p$ for the velocity then one has
\be
\frac{E^2-p^2}{p^2} \,\approx\, 2 \left(\frac{E}{p}-1\right) \,=\, 2 \left({\bar v}(p) -1\right),
\ee
with
\be
{\bar v}(p) \,=\, \frac{1}{p} \int_0^p \text{d}k\, v(k)\,.
\ee

The neutrino decay width can then be rewritten in terms of the velocity of the decaying neutrino as
\be
\Gamma_\nu \sim \frac{3G_F^2}{4\pi^3}\, E_\nu^5 \,({\bar v}(p)-1)^4 \,.
\label{Gamma}
\ee

All the discussion from now on will be based on this relation between the neutrino decay width and the difference of the neutrino velocity with respect to the speed of light.

\section{Phenomenological consistency of unstable superluminal neutrinos in OPERA and SN1987A}

From the previous result we can estimate the mean free path of superluminal neutrinos as
\be
\l_\nu = \Gamma_\nu^{-1} \sim 6 \times \left(\frac{100\,\text{MeV}}{E_\nu}\right)^5 ({\bar v}(p)-1)^{-4}\,\,\text{m}.
\label{rlm}
\ee
Since this expression depends on $\bar{v} (p)$ instead of $v(p)$, we cannot directly apply the OPERA result or the SN1987A bound for the neutrino velocity to obtain constraints derived from neutrino decay. In order to go further one needs to specify the difference of the neutrino velocity with respect to the speed of light as a function of the momentum $p$, at least up to the highest momenta for which we would like to apply Eq.~(\ref{rlm}).

The simplest possibility is to consider a momentum independent velocity of propagation. But in this case there is no neutrino splitting since the kinematic discussion based on energy-momentum conservation reduces to the SR analysis. In fact the expression~(\ref{Gamma}) does not reproduce this result because the proportionality between the typical squared transverse momenta of the neutrinos in the final state and the difference of the square of the energy and the square of the momentum of the decaying neutrino ($p_\perp^2 \propto (E^2 - p^2)$), which is the main approximation used in the derivation of the result~(\ref{Gamma0}), is not justified. However the stringent limit on $v-1$ from SN1987A excludes the interpretation of the OPERA result as due to a superluminal momentum independent velocity.

The next simplest possibility to consider is a momentum dependent power law for the difference of the velocity of neutrinos and the speed of light
\be
v(p) - 1 \,=\, \left(\frac{p}{\Lambda}\right)^n,
\label{npowerlaw}
\ee
where $\Lambda$ is the high-energy scale associated to new physics, $p\ll \Lambda$.
In this case one has
\be
{\bar v}(p) - 1 \,=\,  \frac{v(p) -1}{n+1}
\label{vbar-v}
\ee
and
\be
l_\nu \sim 6\, (n+1)^4 \times \left(\frac{100\,\text{MeV}}{E_\nu}\right)^5 (v(p)-1)^{-4}\,\,\text{m}.
\label{lnu}
\ee
Let us first examine the compatibility between the OPERA and the SN1987A results assuming the power-law behavior Eq.~(\ref{npowerlaw}) up to the energies of the CNGS neutrino beam used in the OPERA experiment ($\langle E_\nu \rangle=17\,\text{GeV}$~\cite{OPERA}).

For $n=1$, taking into account that the energies of the neutrinos from SN1987A are three orders of magnitude smaller than the energy of neutrinos in the OPERA analysis, then the bound on the velocity of propagation for neutrinos in OPERA inferred from supernova (SN) observations is $v-1 \lesssim 2 \times 10^{-6}$, excluding once more the interpretation of the OPERA result as due to superluminal velocities.

In the case of a quadratic momentum dependence ($n=2$) a velocity excess with respect to the speed of light of $2.5 \times 10^{-5}$ at OPERA leads to $v-1 \sim 2.5 \times 10^{-11}$ at supernova energies and a difference of time of flight of neutrinos with respect to gamma rays of around 2 minutes. This implies a spread in the time of arrival of neutrinos of the same order, which is only very marginally compatible with the detection of neutrinos in a 15 second interval~\cite{Giunti}. In any case, we see that a slightly steeper decrease in the velocity from OPERA energies to supernova energies would make the OPERA interpretation of superluminal neutrinos compatible with SN1987A observations.

Let us now accept a situation in which the OPERA and SN1987A results are compatible and study the implications of Eq.~(\ref{rlm}) considering the simple power-law behavior Eq.~(\ref{npowerlaw}). In the case of supernova neutrinos, taking a typical energy of $E_\nu\sim (10$--100)\,\text{MeV} and the maximum velocity of propagation consistent with SN1987A, Eq.~(\ref{lnu}) gives a decay length for the neutrino much larger than the size of the observable universe. However, the mean free path of a neutrino with an energy of 20\,GeV and a velocity $(1+2.5 \times 10^{-5})$ as observed by OPERA is $l_\nu \sim 5\,(n+1)^4\times 10^4\,\text{km}$, which implies the impossibility of the observation of cosmological neutrinos of this energy.

In fact we can use the relation~(\ref{lnu}) to put limits on the velocity from different observations. We have
\be
\left(\frac{v-1}{n+1}\right)^4 \sim 6 \times \left(\frac{\text{m}}{l_\nu}\right) \left(\frac{100\,\text{MeV}}{E_\nu}\right)^5,
\ee
and then lower bounds on the mean free path inferred from the detection of neutrinos of a given energy from a distant source can be translated into upper bounds on the velocity of superluminal neutrinos.
In the case of supernova neutrinos, for an energy of $E=20\,$MeV, the bound $l_\nu>50\,$kpc gives $(v-1)\lesssim 6\,(n+1)\times 10^{-5}$, which is less stringent than the one obtained from the time of arrival of neutrinos, in agreement with what is said in the previous paragraph. In the case of the OPERA experiment, the bound $l_\nu>730\,$km (the distance between the emission at CERN and the detection of the neutrinos at Gran Sasso), gives $(v-1)\lesssim 7\,(n+1)\times 10^{-5}$, so that the simple observation of the expected flux at OPERA puts an upper bound on the superluminal velocities that can be measured in such an experiment which is only slightly above the reported value.

In conclusion, the OPERA result for a superluminal velocity of neutrinos at energies around $20\,$GeV and the bound obtained from SN1987A at an energy three orders of magnitude lower are not incompatible with the constraints coming from the process of neutrino splitting. However, if one takes seriously the OPERA result, these constraints will have deep implications for neutrino physics at a higher energy, as we will now examine.

\section{Consequences of superluminal velocities in neutrino oscillations and cosmological neutrinos}

The decay of superluminal neutrinos may have important consequences in neutrino oscillations and the observation of cosmological neutrinos if one accepts the validity of the OPERA measurement. From the results obtained above, it is clear that neutrino splitting does not play any role in solar neutrino experiments, which involve energies of the same order as for supernova neutrinos. In order to consider the implications for atmospheric neutrino oscillations, we need again to make some assumption on the energy behavior of $v(p)$ at energies around and higher than the $10\,$GeV scale.

In the previous Section we saw how the OPERA result can be made consistent with the observations from SN1987A if one assumes a dependence of the velocity of neutrinos which grows quadratically or quicker with the neutrino momentum between both energy ranges. However, OPERA also reports an independence of the velocity in the energy domain explored by the experiment, within the statistical accuracy of the measurement~\cite{OPERA}. The consistency of this result with SN1987A makes therefore necessary a change in the momentum dependence at this energy domain.

As an example, let us consider a simple model for $v(p)$ satisfying both the compatibility between OPERA and SN1987A, and the independence in energy at the OPERA energy domain. We consider a ``low-momentum'' dependence for the velocity
\be
v(p) - 1 \,=\, (v_0 -1) \left(\frac{p}{\Lambda_0}\right)^{n_<},
\ee
with $n_<$ an exponent which has to be slightly bigger than 2 covering the momentum interval between OPERA and SN1987A. This is combined with a ``high-momentum'' dependence
\be
v(p) - 1 \,=\, (v_0 -1) \left(\frac{\Lambda_0}{p}\right)^{n_>},
\ee
which induces a suppression of the deviations from SR at high energies implemented by an exponent $n_>$ trying to escape to contradictions with high energy neutrino observations. The scale $\Lambda_0$ which parametrizes the turning point from an increase to a decrease of the neutrino velocity should be in the range of the OPERA recent observations in order to reproduce the values for the velocities in the high and low energy bins~\cite{OPERA}.

With the previous momentum dependence one can calculate
\be
\bar{v}(p)-1 \,=\, (v_0-1) \left[N\, \frac{\Lambda_0}{p} - \frac{1}{n_>-1} \left(\frac{\Lambda_0}{p}\right)^{n_>}\right],
\ee
where
\be
N \,=\, \frac{n_>}{n_>-1}-\frac{n_<}{n_<+1}\,\,,
\ee
and the mean free path of ``high-momentum'' neutrinos which results by inserting in Eq.~(\ref{rlm}) the result of $\bar{v}(p)-1$ in this simple model. In the case $p\gg \Lambda_0$ one has
\be
\l_\nu \sim 5 \cdot 10^7 \times \left(\frac{20\,\text{GeV}}{E_\nu}\right)^5
\left(\frac{(v_0-1)}{2.5 \times 10^{-5}}\right)^{-4} \left(\frac{E_\nu}{\Lambda_0}\right)^4 N^{-4} \,\text{m}.
\label{rlm2}
\ee

Eq.~(\ref{rlm2}) gives then a decay length of the order of the Earth diameter for neutrino energies of the order of 100\,GeV. One should expect then  a clear suppression in the high energy tail of the downside atmospheric neutrino spectrum with respect to the upside neutrino spectrum as a consequence of the splitting of superluminal neutrinos. Alternatively, the absence of such suppression would cast serious doubts on the validity of the OPERA result.

The consequences for cosmological neutrinos depend once more on the evolution of $v(p)$ well above the energies explored by OPERA, but in general Eq.~(\ref{rlm2}) also implies the absence of ``high-momentum'' neutrinos  coming from galactic and extragalactic sources. The nonobservation of a diffuse neutrino flux on top of the high energy atmospheric neutrino flux by IceCube~\cite{icecube} could be due to the cutoff induced by neutrino splitting. On the other hand a future observation of extragalactic high energy neutrinos will also give very strong constraints on superluminal velocities for neutrinos.

\section{Concluding remarks}

A consequence of superluminal neutrinos is the possibility of neutrino decay. We have considered the phenomenological constraints of this fact in relation with the recent claim of observation of superluminal neutrinos by OPERA. The OPERA result and the SN1987A bound can be made compatible in the context of departures of SR of the form
\be
v - 1 \,\sim\, \left(\frac{p}{\Lambda}\right)^{n_<}
\ee
with
\be
\Lambda \,=\, \Lambda_0 \left(v_0 -1\right)^{-1/n_{<}}.
\ee
With $n_< =2$, $\Lambda_0 = 20\,\text{GeV}$ and $v_0-1=2.5\times 10^{-5}$ one has a scale $\Lambda$ for the deviations from SR of $4\,\text{TeV}$, many orders of magnitude below the Plank scale which is the natural scale for Lorentz violating effects induced by quantum gravity. It is this aspect of the recent claim of OPERA which makes so difficult to find a framework consistent with all observations of neutrinos. The simple model considered in this work is a purely phenomenological attempt in this direction.

We have estimated the decay length of neutrinos for neutrino splitting as a function of their velocity. This allows us to study the consistency of unstable neutrinos with the OPERA and SN1987A results. 
While both observations can be made consistent with the possibility of neutrino decay, we find that a value just one order of magnitude larger for $v-1$ of superluminal neutrinos than the value reported by OPERA would make them decay before arriving to their detector.

In our study of neutrino splitting we have neglected neutrino masses and the corresponding threshold for the reaction. Such a threshold will be $p_{\text{th}} \sim \sqrt{m_\nu \Lambda}$ which is of the order of $\text{MeV}$, much below the energies considered in the neutrino splitting analysis. This justifies the approximation to neglect neutrino masses.

The decay of superluminal neutrinos may also be relevant in the observation of atmospheric neutrinos and cosmological neutrinos. We have shown that the high-energy tail of the atmospheric neutrino spectrum is affected by superluminal neutrino splitting. The OPERA result also implies quite generally the impossibility to observe cosmological neutrinos.

Our analysis rely on the validity of Eq.~(\ref{Gamma}) for the decay width in neutrino splitting. A limitation of that formula is that the approximations made to derive it are not valid in the case of a momentum independent velocity. This does not affect the analysis made for energies lower than the OPERA range, since the compatibility of OPERA with SN1987A imposes an energy dependent velocity for which Eq.~(\ref{Gamma}) is correct. However, the OPERA data suggest a constant value of the velocity at this energy range. If this were a true effect going on for larger energies, Eq.~(\ref{Gamma}) would loose its validity and the analysis for high energies would change.

While we were completing this manuscript two new papers appeared pointing out the possibility to put constraints on superluminal neutrinos from neutrino decay~\cite{cohen,bi}. Our conclusions differ from those of Ref.~\cite{cohen} due to the different formulae for the decay width. While we know that our expression requires a momentum dependence of the velocity justifying the approximations on the angular dependence of the decay, it looses its validity for a near independent velocity. In Ref.~\cite{bi} the relevance of departures from SR in the reactions responsible for the production of neutrinos is proposed as a complementary source of constraints together with the decay of neutrinos. A generalization of these analysis including a general dependence on momenta for the superluminal velocity of neutrinos should be considered.

As a final word of caution, the present analysis (and those of Refs.~\cite{cohen,bi}) rely on the use of the standard energy-momentum conservation law, which converts a forbidden reaction in SR into an allowed one for superluminal neutrinos. There are however theoretical scenarios, such as Doubly Special Relativity (DSR) theories~\cite{DSR}, which consider departures from SR while maintaining a relativity principle, requiring then a modification of the conservation laws. Quite generally, the consistency between the conservation laws of the energy momentum, the dispersion relation, and the relativity principle produces in these frameworks cancellations which tend to hold the SR results of the kinematical analysis of a process~\cite{ASR,DSR3}. This could even forbid the decay of superluminal neutrinos, depending on the specific DSR model.


\section*{Acknowledgments}
This work is supported by CICYT (grant FPA2009-09638) and DGIID-DGA (grant
2010-E24/2).



\end{document}